\def\den{\hbox{den}}

\def\tr{\hbox{tr}}
\def\ln{\ell{n}}
\documentstyle[a4,12pt]{article}
\begin{document}
\begin{titlepage} \vspace{0.2in} \begin{flushright}
MITH-95/16 \\ \end{flushright} \vspace*{1.5cm}
\begin{center} {\LARGE \bf  A Lattice Chiral Gauge Theory
with Multifermion Couplings
\\} \vspace*{0.8cm}
{\bf She-Sheng Xue$^{(a)}$}\\ \vspace*{1cm}
INFN - Section of Milan, Via Celoria 16, Milan, Italy\\ \vspace*{1.8cm}
{\bf   Abstract  \\ } \end{center} \indent

Analyzing an $SU_L(2)$-chiral gauge theory with external multifermion
couplings,
we find a possible scaling region
where doublers decouple by acquiring chiral-invariant masses and $\psi_R$ is
free mode owing to the $\psi_R$-shift-symmetry, the chiral continuum theory
of $\psi_L$ can be defined. This is not in agreement
with the general belief of the failure of theories so constructed .

\vfill \begin{flushleft}  September, 1995   \vspace*{3cm} \\
\noindent{\rule[-.3cm]{5cm}{.02cm}} \\
\vspace*{0.2cm} \hspace*{0.5cm} ${}^{a)}$
E-mail address: xue@milano.infn.it\end{flushleft} \end{titlepage}

\noindent
{\bf 1.}\hspace*{0.3cm}
It is a long standing problem to regularize chiral gauge theories (CGT)
on a lattice and it seems that none of the methods proposed has been
consistently and completely
demonstrated both to ensure that an asymptotically chiral gauge theory in the
continuum limit really exists and to provide a framework for doing
nonperturbative calculation in these theories \cite{p}. It is generally
believed that the constructions \cite{ep,xue} of CGT on the lattice with
external multifermion couplings fail to give chiral gauged fermions in the
continuum limit for the reason\cite{gpr} that the theories so constructed
undergo spontaneous symmetry breaking and their phase structure is similar
to that of the Smit-Swift model\cite{ss}, which has been very carefully studied
and shown to fail. Nevertheless, we believe that further considerations of
constructing CGT on a lattice with external multifermion couplings and careful
studies of the spectrum in each phase of such a constructed theory are
necessary.
In fact, we find the possible scaling region of defining continuum chiral
fermion in
such a formulation of CGT on the lattice.

Let us consider the following fermion action of the $SU_L(2)$ CGT on a lattice
with two external multifermion couplings.
\begin{eqnarray}
S&\!=\!&{i\over 2a}\sum_x\left(\bar\psi^i_L(x)\gamma_\mu D^\mu_{ij}\psi^j_L(x)+
\bar\psi_R(x)\gamma_\mu\partial^\mu\psi_R(x)\right)\label{action}
\\
&\!+\!&
\sum_x\!\left(g_1\!\bar\psi^i_L(x)\!\cdot\!\psi_R(x)\bar\psi_R(x)\!\cdot\!\psi_L^i(x)
\!+\!g_2\!\bar\psi^i_L(x)\!\cdot\!\partial^2\psi_R(x)
\partial^2\bar\psi_R(x)\!\cdot\!\psi_L^i(x)\right),\nonumber
\end{eqnarray}
where ``$a$'' is the lattice spacing;
$\psi^i_L$ ($i=1,2$) is an $SU_L(2)$ gauged doublet, $\psi_R$ is an $SU_L(2)$
singlet\footnote{$\psi_R$ is treated as a ``spectator'' fermion since we do not
consider $U_Y(1)$ symmetry.} and both are two-component Weyl fermions.
The second multifermion coupling $g_2$, where $\partial^2\psi_R(x)=\sum_\mu
\left[ \psi_R(x+\mu)+\psi_R(x-\mu)-2\psi_R(x)\right]$, is a dimension-10
operator relevant only for doublers $p=\tilde p+\pi_A, (\tilde p\simeq 0$
and $\pi_A$ runs over fifteen lattice momenta $\pi_A\not=0$.),
but irrelevant for normal modes $p=\tilde p$ of the $\psi^i_L$ and
$\psi_R$. In addition to the exact local $SU(2)$ chiral
gauge symmetry and the global flavour symmetry $SU_L(2)\otimes U_R(1)$,
the action (\ref{action}) possesses a
$\psi_R$-shift-symmetry\cite{gp}, $\psi_R(x) \rightarrow \psi_R(x)+{\rm
const.},
$ when $g_1=0$. The Ward identity corresponding to this $\psi_R$-shift-symmetry
is
\begin{equation}
{i\over 2a}\gamma_\mu\partial^\mu\psi^\prime_R(x)
+g_1\langle\bar\psi^i_L(x)\!\cdot\psi_R(x)\psi_L^i(x)\rangle
\!+g_2\!\langle\partial^2\!\left(\bar\psi^i_L(x)\!\cdot\!
\partial^2\psi_R(x)\psi_L^i(x)\right)\rangle+\Gamma^{(1)}_{\bar R}(x)=0,
\label{w}
\end{equation}
where ``prime'' fields $\phi^\prime_k(x)\sim {\delta\ln
W\over\delta\eta_k(x)}$,
the one-particle irreducible
vertex functions $\Gamma(\phi^\prime_k)$ are the Legendre transformation
of the generating functional $W=-\ln Z(\eta_k)$ and $\Gamma^{(1)}_{\bar R}(x)=
{\delta\Gamma\over\delta\bar\psi^\prime_R(x)}$.
$\langle\cdot\cdot\cdot\rangle$ is an expectation value with respect to the
partition function $Z(\eta)$. Based on this Ward
identity (\ref{w}), one can get all one-particle irreducible vertices
containing
at least one external $\psi_R$. $\Gamma^{(2)}_{R\bar R}(p)={i\over a}
\gamma_\mu\sin p^\mu P_R$(the $\psi_R$ is free field\cite{gp}),
$\Gamma^{(2)}_{L\bar R}(p)=\Sigma^i(p)\simeq a^2(g_1+2g_2w(p))
\langle\bar\psi^i_L\psi_R\rangle$, where $w(p)=\sum_\mu (\cos p_\mu-1)$, and
the four-point vertex function is given by

\vskip0.5cm
\begin{equation}
\hskip3cm =\Gamma^{(4)}_{L\bar L R\bar R}
(p,p',q)=g_1+4g_2w(p+{q\over 2})w(p'+{q\over 2}),
\label{v}
\end{equation}
where $p+{q\over 2}$ and $
p'+{q\over 2}$ are the momenta of the external $\psi_R$. All other one-particle
irreducible vertices $\Gamma^{(3)}_{L\bar L \bar R}=\Gamma^{(3)}_{L R\bar R}=
\Gamma^{(n)}_{\cdot\cdot\cdot\bar R}=0 (n>4)$ identically. We find that when
$g_1=0$, the $\Gamma^{(2)}_{L\bar R}$ and $\Gamma^{(4)}_{L\bar L R\bar R}$ for
the normal mode of the $\psi_R$ are vanishing at least $O((ma)^2)$, where $m$
is the scale of the continuum limit.  This indicates
that when $g_1=0$, the normal mode of the $\psi_R$ completely decouples and
dose not form any bound states with other modes.

\vskip0.7cm
\noindent
{\bf 2.}\hspace*{0.3cm}
Our goal is to seek a possible regime, where an undoubled $SU(2)$-chiral
gauged fermion content is exhibited in the continuum limit
in the phase space $(g_1,g_2,g)$,
where ``$g$'' is the gauge coupling,  regarded to be a truly small
perturbation $g\rightarrow 0$ at the scale of the continuum limit we consider.
In the weak coupling limit, $g_1\ll 1$ and $g_2\ll 1$ (indicated 1 in fig.1),
the action (\ref{action}) defines an $SU_L(2)\otimes U_R(1)$ chiral continuum
theory with a doubled and weakly interacting fermion spectrum that is not the
continuum theory we seek.

Let us consider the phase of spontaneous symmetry breaking in the weak-coupling
$g_1,g_2$ limit. Based on the analysis of large-$N_f$ ($N_f$ is an
extra fermion index, e.g.,~color, $N_c$) weak coupling expansion, we
show that the multifermion couplings in (\ref{action}) undergo Nambu-Jona
Lasinio (NJL) spontaneous chiral-symmetry breaking\cite{njl}. In this symmetry
breaking phase (indicated 2 in fig.1) the
$\psi^i_L$ and $\psi_R$ in (\ref{action}) pair up to be a massive Dirac fermion
violating
$SU(2)$-chiral symmetry. The spectrum of the theory (\ref{action}) can be
written
as,
\begin{equation}
S^{-1}_b(p)=\left(\matrix{&P_R{i\over a}\sum_\mu\gamma_\mu f^\mu_L(p)^{ij}P_L&
P_R\Sigma^j(p)P_R\cr
&P_L\Sigma^i(p)P_L&P_R{i\over a}\sum_\mu\gamma_\mu \sin p^\mu P_L}\right),
\label{sb}
\end{equation}
where the fermion self-energy function is given by ($N_f\rightarrow\infty$)
\begin{equation}
\Sigma^i(p)=4\int_q{\Sigma^i(q)\over\den(q)}\left(\tilde g_1+4\tilde
g_2w(p)w(q)
\right)
\label{se}
\end{equation}
where $\int_q\equiv \int_\pi^\pi{d^4q\over (2\pi)^4}, \den(q)\equiv
\sum_\rho\sin^2q_\rho +(\Sigma^i(q)a)^2$ and $\tilde g_1\equiv g_1N_fa^2,
\tilde
g_2\equiv g_2N_fa^2$. Using the paramatrization $\Sigma^i(p)=\Sigma^i(0)+\tilde
g_2 v^iw(p)$ and $\Sigma^i(0) =\rho v^i$\cite{gpr}, where $\rho$ depends only
on
couplings $\tilde g_1, \tilde g_2$, and $v^i$ plays a role as the v.e.v.
violating $SU(2)$-chiral symmetry, we can solve the gap-equation (\ref{se}).
For $v^i=O({1\over a})$, one obtains
\begin{equation} \rho={\tilde g_1\tilde
g_2 I_1\over 1-\tilde g_1 I_\circ}; \hskip1cm \rho={1-4\tilde g_2 I_2\over
4 I_1},\label{rho}
\end{equation} where the functions
$I_n=4\int_q{w^n(q)\over\den(q)}$. Eq.(\ref{rho}) leads to $\tilde g_1=0,
\rho=0$ and $\Sigma^i(0)=0$, this result is due to the Ward identity (\ref{w}).
This means that on the line $g_1$=0,
normal modes $p=\tilde p\simeq 0$ of the $\psi^i_L$ and $\psi_R$ are
massless and their 15 doublers $p=\tilde p+\pi_A$ acquire chiral-variant masses
$\Sigma^i(p)$ through the multifermion coupling $g_2$ only
($1-4\tilde g_2 I_2=0$).

As for the function $f^\mu_L(p)^{ij}$ in eq.(\ref{sb}),
it depends on
the dynamics of the left-handed Weyl fermion $\psi_L^i$ in this
region. In large-$N_f$
calculation at weak couplings we are able to evaluate the function
$f^\mu_L(p)^{ij}=\delta_{ij}\sin p_\mu Z_2(p)$ (see fig.2) and the wave
function renormalization is given by
\begin{eqnarray}
Z^{-1}_2(p)\!&=&\!1\!-\!{2\over N_f}\!\int_{k,l}\!\left(\tilde g_1\!+\!4
\tilde g_2w(k)w(k\!-\!{l\over2})\right)^2\!
{\sum_{\mu\nu}\!\gamma_\mu\gamma_\nu
\sin(p\!-\!k)^\mu\sin p^\nu\over\sum_{\lambda\rho}\sin^2(p-k)_\lambda
\sin^2 p_\rho}R(k,\!l)\nonumber\\
R(k, l)&=&{\sum_{\sigma}\sin(k-{l\over2})^\sigma\sin (k+
{l\over2})_\sigma\over\sum_{\sigma\sigma^\prime}
\sin^2(k-{l\over2})^\sigma\sin^2(k+
{l\over2})^{\sigma^\prime}}.
\label{z2}
\end{eqnarray}
Assuming the symmetry breaking takes place in the direction 1 in the
2-dimensional space of the $SU(2)$-chiral symmetry ($v^1\not=0, v^2=0$), we
find the
following fermion spectrum, containing a doubled Weyl fermion $\psi^2_L$ and a
undoubled Dirac fermion made by the Weyl fermions $\psi^1_L$ and
$\psi_R$,
\begin{eqnarray}
S_{b1}^{-1}(p)&=&{i\over a}\sum_\mu\gamma_\mu\sin p_\mu Z_2(p)P_L
+{i\over a}\sum_\mu\gamma_\mu \sin p^\mu P_R+v^1(\rho+\tilde
g_2w(p))\label{sb1}
\\
S_{b2}^{-1}(p)&=&{i\over a}\sum_\mu\gamma_\mu\sin p_\mu Z_2(p)P_L.
\label{sb2}
\end{eqnarray}
The $SU_L(2)\otimes U_R(1)$ chiral symmetry is realized to be
$U_L(1)\otimes U(1)$
with three Goldstone modes and a massive Higgs mode that are not presented in
this short report.
As $v^1\rightarrow 0$, eq.(\ref{rho}) gives a critical line $\tilde g^c_1
(\tilde g^c_2)$ of characterizing NJL spontaneous chiral symmetry breaking
and $\tilde g^c_1=0.4, \tilde g^c_2=0$; $\tilde g^c_1=0,
\tilde g^c_2=0.0055$ (indicated 2 in fig.1).
These critical values are sufficiently small evenfor $N_f=1$.

This broken phase cannot be a candidate for a real chiral
gauge theory (e.g.,~the Standard Model) for the reasons that $(i)$ $\psi^2_L$
is
doubled (\ref{sb2}); $(ii)$ the spontaneous symmetry breakdown of the
$SU_L(2)$-chiral symmetry is caused by the hard breaking Wilson
term\cite{wilson} (\ref{sb1})(dimension-5 operator), which must contribute
the intermediate gauge boson masses through the perturbative
gauge interaction and disposal of Goldstone modes. The intermediate gauge boson
masses turn out to be $O({1\over a})$. This, however, is
phenomenologically unacceptable.

\vskip0.7cm
\noindent
{\bf 3.}\hspace*{0.3cm}
We turn to the strong coupling region, where $g_1(g_2)$ is sufficiently larger
than a certain critical value $g_1^c(g_2^c)$ (indicated 3 in fig.1).
Analogously to the analysis and
discussions
of Eichten and Preskill (EP) \cite{ep}, we can show that the
$\psi^i_L$ and $\psi_R$ in (\ref{action}) are bound up to form the composite
Weyl fermions $(\bar\psi^i_L\cdot\psi_R)\psi^i_L$ (left-handed
$SU_L(2)$-neutral) and $(\bar\psi_R\cdot\psi^i_L)\psi_R$ (right-handed
$SU_L(2)$-charged) and these bound fermion states respectively pair up with the
$\bar\psi_R$ and $\bar\psi_L^i$ to be massive, neutral $\Psi_n$ and charged
$\Psi_c^i$ Dirac modes consistently with the $SU_L(2)\otimes U_R(1)$ chiral
symmetry.

The second multifermion coupling $4g_2w(p+{q\over2})w(p'+{q\over2})$ in
(\ref{v}
) gives different
contributions to the effective value of $g_1$ at large distance for sixteen
modes of the $\psi_L^i$
and $\psi_R$ in the action (\ref{action}). Let us consider the multifermion
couplings of each mode ``$p$'' of the $\psi_L^i$
and $\psi_R$, namely, we set $p=p', q=\tilde q\ll 1$ in the
four-point vertex (\ref{v}). Using strong multifermion coupling,
$\Gamma^{(4)}_{L\bar L R\bar R}=g_1+4g_2w^2(p)\gg 1$, expansion\footnote{
This is just a hopping parameter expansion.} and recursion relation \cite{we}
for each mode ``$p$'' of the $\psi^i_L$ and $\psi_R$,
in the lowest nontrivial order we calculate the propagators of neutral and
charged Dirac modes to be
\begin{eqnarray}
S_n(p)\!&\simeq\!&{{i\over a}\sum_\mu\gamma_\mu\sin p^\mu +M\over
{1\over a^2}\sum_\rho\sin^2p_\rho+M^2}
\label{sn1}\\
S_c(p)_{ij}\!&\simeq\!&\delta_{ij}{{i\over a}\sum_\mu\gamma_\mu\sin p^\mu +M
\over {1\over a^2}\sum_\rho\sin^2p_\rho+M^2},
\label{sc1}
\end{eqnarray}
where the chiral-invariant masses $M^2=16a^2(g_1+4g_2w^2(p))^2$ and the
spectrum
is vector-like.

The critical value $g^c_1(g^c_2)$ can be determined by considering \cite{gpr}
the propagators $G^{ij}_{1,2}(q)$ of four composite scalars
$A_1^i={1\over\sqrt{2}}(\bar\psi^i_L\!\cdot\!\psi_R+\bar\psi_R\!\cdot\!\psi^i_L)$
and
$A_2^i={i\over\sqrt{2}}(\bar\psi^i_L\!\cdot\!\psi_R-\bar\psi_R\!\cdot\!\psi^i_L)$,
which are the real and imaginary parts of a complex composite field
${\cal A}^i=\bar\psi_R\cdot\psi^i_L$. Again using the strong coupling,
$\Gamma^{(4)}_{L\bar L R\bar R}=g_1+4g_2w^2(p)\gg 1$,
expansion and recursion relation \cite{we}
for each mode ``$p$'' of the $\psi^i_L$ and $\psi_R$, in the lowest nontrivial
order we find
these four massive composite scalar modes,
\begin{equation}
G^{ij}_{1,2}(\tilde q)\simeq {\delta_{ij}\over {4\over
a^2}\sum_\mu\sin^2{\tilde
q_\mu\over 2}
+\mu^2};\hskip0.5cm \mu^2\simeq 16\left(g_1+4g_2w^2(p)-{1\over2 a^2}\right),
\label{mas}
\end{equation}
which are degenerate owing to the
exact $SU(2)$-chiral symmetry. A
spontaneous symmetry breaking $SU(2)\rightarrow U(1)$ occurs, where $\mu^2>0$
turns to $\mu^2<0$. Eq.(\ref{mas}) for $\mu^2=0$ gives rise to the critical
lines:  $g_1^ca^2=0.5, g_2=0$; $g_1=0, a^2g_2^{c,b}=0.002$ where the first
binding threshold of the doubler $p=(\pi,\pi,\pi,\pi)$ is, and $g_1=0,
a^2g_2^{c,a}=0.031$ where the last binding threshold of the doublers
$p=(\pi,0,0,0)$ is, inbetween (indicated 4 in fig.1) there are the binding
thresholds
of the doublers $p=(\pi,\pi,0,0)$ and $p=(\pi,\pi,\pi,0)$, and the binding
thresholds of the different doublers $p\not=p'$, which can be analogously
calculated.
Above $g_1^{c,a}$ {\it all} doublers are supposed to be
bound, as indicated 5 in fig.1. As for the normal modes of the $\psi^i_L$
and $\psi_R$, when $g_1\ll 1$, the multifermion coupling,
$\Gamma^{(4)}_{L\bar L R\bar R}=g_1+4g_2w^2(\tilde p)$, is no longer strong
enough to form the bound states $(\bar\psi^i_L\cdot\psi_R)\psi^i_L$,
$(\bar\psi_R\cdot\psi^i_L)\psi_R$ and ${\cal A}^i$ unless $a^2g_2
\rightarrow\infty$. It is conceivable that the critical line for normal modes
$g_1+ag_2O((ma)^4)-{1\over 2a^2}=0$ analytically continues to the limit
$g_2^{c,\infty}=g_1\rightarrow 0,g_2\rightarrow \infty$.

Thus, as expected in ref.\cite{ep}, several wedges open up as $g_1, g_2$
increase in the NJL phase (indicated 5 in fig.1), inbetween the critical lines
along which bound states of normal modes and doublers of the $\psi^i_L$ and
$\psi_R$ respectively approach their thresholds. In the initial part of the NJL
phase, the normal modes and doublers of the $\psi^i_L$ and $\psi_R$ undergo the
NJL phenomenon and contribute to eq.(\ref{sb}) as discussed in section 2. As
$g_1,g_2$ increase, all these modes, one after another, gradually disassociate
from the NJL phenomenon and no longer contribute to eqs.(\ref{sb}). Instead,
they turn to associate with the EP phenomenon and contribute to
eqs.(\ref{sn1},\ref{sc1}).  The first and last doublers of the $\psi^i_L$ and
$\psi_R$ making this transition are $p=(\pi,\pi,\pi,\pi)$ and
$p=(\pi,0,0,0)$ respectively. At the end of this sequence, normal
modes ($p=\tilde p)$ make this transition, due to the fact that they possess
the different effective multifermion coupling
$\Gamma^{(4)}_{L\bar L R\bar R}=g_1+4g_2w^2(p)$.

Had these critical lines separated the two symmetric phases, (strong couplings
and the weak coupling symmetric phases) we would have found a threshold over
which all doublers of the $\psi_L^i$ and $\psi_R$ decouple by acquiring chiral
invariant masses (\ref{sn1},\ref{sc1}) and normal modes of the $\psi_L^i$ and
$\psi_R$ remain massless and free, and we might obtain a theory of massless
free
chiral fermions \cite{ep}. However, this is not real case \cite{gpr}. As has
been seen in eq.(\ref{mas}), turning $\mu^2>0$ to $\mu^2<0$ indicates a phase
transition between the strong coupling symmetric phase to the spontaneous
chiral symmetry breaking phase, which separates the strong coupling and
weak coupling symmetric phases.

The possible resolution of this undesired situation is that we find a wedge in
which the doublers of the $\psi^i_L$ and $\psi_R$ have formed bound states
$(\bar\psi_R\cdot\psi^i_L)\psi_R$ and $(\bar\psi^i_L\cdot\psi_R)\psi^i_L$ via
the EP phenomenon, while the normal modes of the $\psi^i_L$ and
$\psi_R$ have neither formed such bound states yet and nor are they
associated with the NJL-phenomenon.

Within the last wedge (indicated 5 in fig.1) between two the thresholds
$g_2^{c,a}$ and $g_2^{c,\infty}$, all doublers of the $\psi^i_L$ and $\psi_R$
acquire chiral-invariant masses and decouple (considering
eqs.(\ref{sn1},\ref{sc1}) as the propagators for doublers $p=\tilde p+\pi_A$
{\it only}) and we
have the undoubled low-energy spectrum that involves only the normal modes of
the $\psi^i_L$ and $\psi_R$. However, because of the multifermion coupling
$g_1\not= 0$, these normal modes of $\psi^i_L$ and $\psi_R$ still
remain in the NJL broken phase, the $SU(2)$-chiral symmetry is violated by
$\Sigma^1(0)=\rho v^1$, to which only normal modes contribute. The propagators
of the normal modes in this wedge should be the same as
eqs.(\ref{sb1},\ref{sb2})
for $p=\tilde p$. However, when $g_1\not= 0$, the normal mode of the $\psi_R$
is not guaranteed to completely decouple from that of the $\psi^i_L$.

Once we go onto the line A ($g_1=0, g_2^{c,a}<g_2<g_2^{c,\infty}$ as indicated
in fig.1), the spectrum is undoubled for $g_2>g_2^{ca}$ and as the results of
the $\psi_R$-shift-symmetry of the action (\ref{action}): (i) the normal mode
of the $\psi_R$ is a free mode; (ii) the interacting vertex
$\Gamma^{(4)}_{L\bar L R\bar R}=4g_2w^2(\tilde p)\ll 1$ for the normal modes,
which prevent the normal modes of the $\psi^i_L$ and $\psi_R$ from binding up
bound states $(\bar\psi^i_L\cdot\psi_R)\psi^i_L$,
$(\bar\psi_R\cdot\psi^i_L)\psi_R$ and ${\cal A}^i$; (iii) the NJL mass term
$\Gamma^{(2)}_{L\bar R}=\Sigma^1(0)=0$ for which the $SU_L(2)\otimes
U_R(1)$-chiral symmetry is
completely restored. In this scaling region, the spectrum consists of the
doublers eq.(\ref{sn1},\ref{sc1}) for $g_1=0$ and $p=\tilde p+\pi_A$, the
massless normal modes eqs.(\ref{sb1},\ref{sb2}) for $g_1=0$ and $p=\tilde p$,
\begin{equation}
S^{-1}_L(\tilde p)^{ij}=i\gamma_\mu\tilde p^\mu\tilde Z_2\delta_{ij}P_L;
\hskip1cm S^{-1}_R(\tilde p)=i\gamma_\mu\tilde p^\mu P_R,
\label{sf}
\end{equation}
which is in agreement with the $SU_L(2)\otimes U_R(1)$ symmetry.
Namely, this normal mode of the $\psi_L^i$ is self-scattering via the
multifermion coupling $g_2$ (see fig.2) without pairing up with any other
modes.
The wave function
renormalization $\tilde Z_2$ can be considered as an interpolating constant
of $Z_2(p)$ eq.(\ref{z2}) for $p=\tilde p\simeq 0$ and $g_1=0$.

In summary, in the scaling region for the long distance, we have the spectrum
that contains the $SU_L(2)$-invariant and $U_R(1)$-covariant neutral Dirac
mode $\Psi_n$
eq.(\ref{sn1})($p\not=\tilde p$); the $U_R(1)$-invariant and
$SU_L(2)$-covariant charged Dirac mode
$\Psi^i_c$ eq.(\ref{sc1})($p\not=\tilde p$); the $U_R(1)$-covariant Weyl mode
$\psi_R$ and the $SU_L(2)$-covariant Weyl mode $\psi^i_L$
eq.(\ref{sf})($p=\tilde p$) as well as the $SU_L(2)\otimes U_R(1)$ covariant
scalar ${\cal A}^i$ eq.(\ref{mas})($q=\tilde q$).
In order to see all possible interactions between
these modes in the scaling region, we consider the one-particle irreducible
vertex functions of these modes. In the light of the exact $SU_L(2)\otimes
U_R(1)$ chiral symmetry and $\psi_R$-shift-symmetry, one can straightforwardlly
obtain non-vanishing vertex functions ($d$=dimensions) at physical momenta
($p=\tilde p, q=\tilde q$): (i) ${\cal A}^j{\cal A}^{j\dagger}{\cal A}^i{\cal
A}^{i\dagger}$ ($d=4$); (ii) $\bar\psi^i_L\psi_L^i{\cal A}^j{\cal
A}^{j\dagger}$, $\bar\Psi^i_c\Psi_c^i{\cal A}^j{\cal A}^{j\dagger}$ and
$\bar\Psi_n\Psi_n{\cal A}^j{\cal A}^{j\dagger}$ ($d=5$), as well as $d>5$
vertex functions. The vertex functions with dimensions $d>4$ vanish in the
scaling region as $O(a^{d-4})$ and we are left with the self-interacting vertex
${\cal A}^j{\cal A}^{j\dagger}{\cal A}^i{\cal A}^{i\dagger}$.

In this scaling region,
the chiral continuum limit is very much like that of
lattice QCD. We need to tune only one coupling $g_1\rightarrow0$ in the
neighborhood of $g_2^{c,a}<g_2<g_2^{c,\infty}$. For $g_1\rightarrow 0$, the
$\psi_R$-shift-symmetry is slightly violated, the normal modes of the
$\psi^i_L$ and $\psi_R$ would couple together to form the chiral symmetry
breaking
term $\Sigma^i(0)\bar\psi^i_L\psi_R$, which is a dimension-3 renormalized
operator and thus irrelevant at the short distance. We desire this scaling
region to be ultra-violet stable, in which the multifermion coupling $g_1$
turns out to be an effective renormalized dimension-4 operator\cite{bar}.

\vskip0.7cm
\noindent
{\bf 4.}\hspace*{0.3cm}
The conclusion of the existence of a possible scaling region for the continuum
chiral theory is plausible and
it is worthwhile to confirm this scenario in different approaches.
However, we are still left with several problems.
Their possible resolutions are mentioned and discussed in this section,
and deserve to be studied in future work.

The question is whether this chiral continuum theory in the scaling region
could
be the correct chiral gauge theory, as the $SU(2)$-chiral gauge coupling $g$
perturbatively is turned on in the theory (\ref{action}). One should expect
a slight change of critical lines (points). We
should be able to re-tune the multifermion couplings ($g_1,g_2$) to compensate
these perturbative changes. In the scaling regime, disregarding those
uninteresting neutral modes, we have the charged modes including
both the $SU(2)$-chiral-gauged, massless normal mode (\ref{sf}) of the
$\psi^i_L$ and the $SU(2)$-vectorial-gauged, massive doublers of the Dirac
fermion
$\Psi^i_c$ (\ref{sc1}), which is made by the 15 doublers of the $\psi^i_L$ and
the 15 doublers of the bound Weyl fermion $(\bar\psi_R\cdot\psi^i_L)\psi_R$.
The gauge
field should not only chirally couple to the massless normal mode of
the $\psi_L^i$ in the low-energy regime, but also vectorially couple to the
massive doublers of Dirac fermion $\Psi_c^i$ in the high-energy
regime. Thus, we expect the coupling vertex of the $SU_L(2)$-gauge field
and the normal mode of the $\psi^i_L$ to be chiral at the continuum
limit.
We are supposed to be able to
demonstrate this point on the basis of the Ward identities associating with the
$SU(2)$-chiral gauge symmetry that is respected by the spectrum in the scaling
regime. In fact, due to the reinstating of the manifest $SU(2)$-chiral gauge
symmetry and corresponding Ward identities of the undoubled spectrum in this
scaling regime, we should then apply the Rome approach\cite{rome} (which is
based
on the conventional wisdom of quantum field theory) to perturbation theory
in the small gauge coupling. It is expected that the Rome approach would work
in the same way but all gauge-variant counterterms are prohibited; the gauge
boson
masses vanish to all orders of gauge coupling perturbation theory for $g_1=0$.

Another important question remaining is how chiral gauge anomalies emerge,
although in this short report the chiral gauge anomaly is cancelled by
purposely choosing
an appropriate fermion representation of the $SU_L(2)$ chiral gauge group.
We know that in the doubled spectrum of naive lattice chiral gauge theory,
the reason for the correct anomaly disappearing in the continuum limit is that
the normal mode and doublers of Weyl fermion produce the same anomaly
these anomalies eliminate themselves
\cite{smit}.
As a consequence of decoupled doublers being given chiral-invariant mass
$(\sim O({1\over a}))$, the survival normal mode of the Weyl fermion
(chiral-gauged,
e.g., $U_L(1)$) should produce the correct anomaly in the continuum limit.
We also have the question of whether the conservation of fermion number would
be violated by the correct anomaly\cite{ep,bank} structure $\tr F\tilde F$ that
is generated by the $SU(2)$ instanton in the continuum limit.

I am grateful to G.~Preparata for advice and discussions and thank M.~Testa,
D.N.~Petcher and M.~Golterman for discussions at Melbourne Lat'95.

\section*{Figure Captions}

\noindent {\bf Figure 1}: \hspace*{0.2cm}
The phase diagram for the theory (\ref{action}) in the $g_1-g_2$
plane (at $g\simeq 0$).

\noindent {\bf Figure 2}: \hspace*{0.2cm}
Contributions to the wave function renormalization $Z_2(p)$ of the $\psi_L^i$.


\begin{thebibliography}{99}
\bibitem{we}
These calculations are similar to that in ref.\cite{ep} and the details will be
reported elsewhere.
\end{thebibliography}
\end{document}